\documentclass[12pt,a4paper]{article}
\usepackage[left=1in,right=1in,top=1.5in,bottom=1.5in]{geometry}

\usepackage{times}
\usepackage{mathrsfs}
\usepackage{amssymb}
\usepackage{amsmath}
\usepackage{ascmac}
\usepackage{amsthm}
\usepackage[dvipdfmx]{graphicx}
\usepackage{booktabs}
\usepackage{setspace}
\usepackage{natbib}
\usepackage{color}
\usepackage{comment}

\usepackage{titlesec}
\titleformat*{\section}{\large\bfseries}
\titleformat*{\subsection}{\it}

\newtheorem{thm}{Theorem}
\newtheorem{algo}{Algorithm}

\def\la{{\lambda}}

\def\th{{\theta}}

\def\bbe{{\text{\boldmath $\beta$}}}
\def\bga{{\text{\boldmath $\gamma$}}}

\def\bla{{\text{\boldmath $\lambda$}}}

\def\bth{{\text{\boldmath $\theta$}}}

\def\bta{{\text{\boldmath $\eta$}}}

\def\thh{{\widehat \th}}

\def\Eh{{\widehat E}}

\def\btah{{\widehat \bta}}

\def\tht{\widetilde{\theta}}
\def\sit{\widetilde{\sigma}}

\def\bb{{\text{\boldmath $b$}}}
\def\bh{{\text{\boldmath $h$}}}

\def\bs{{\text{\boldmath $s$}}}
\def\bu{{\text{\boldmath $u$}}}
\def\bx{{\text{\boldmath $x$}}}
\def\by{{\text{\boldmath $y$}}}
\def\bB{{\text{\boldmath $B$}}}

\def\bU{{\text{\boldmath $U$}}}

\def\bX{{\text{\boldmath $X$}}}
\def\zero{{\text{\boldmath $0$}}}

\title{{\bf Bayesian Benchmarking Small Area Estimation \\
via Entropic Tilting}\footnote{This version: \today}}
\date{}

\begin{document}
\doublespacing
\maketitle

\vspace{-2cm}
\begin{center}
{\large Shonosuke Sugasawa$^1$, Genya Kobayashi$^2$ and Yuki Kawakubo$^3$}
\end{center}

\noindent
$^1$Faculty of Economics, Keio University, Japan\\
$^2$School of Commerce, Meiji University, Japan\\
$^3$Graduate School of Social Sciences, Chiba University, Japan

\vspace{0.5cm}
\begin{center}
{\large\bf Abstract}
\end{center}
Benchmarking estimation and its risk evaluation is a practically important issue in small area estimation. While Bayesian methods have been widely adopted in small area estimation, existing benchmarking approaches are often ad-hoc, such as projecting each MCMC draw to satisfy the constraint. In contrast, our work provides a unified Bayesian formulation based on entropic tilting, which offers a more principled way to define the benchmarked posterior distribution. This approach yields benchmarked point estimates together with coherent uncertainty quantification. We first introduce general Monte Carlo methods for obtaining a benchmarked posterior under hierarchical Bayesian approaches and then show that the benchmarked posterior under empirical Bayesian frameworks can be obtained in an analytical form for some small area models. We demonstrate the usefulness of the proposed method through simulation and empirical studies.

\bigskip\noindent
{\bf Key words}: Markov chain Monte Carlo; posterior distribution; small area estimation

\newpage
\section{Introduction}
Small area estimation is the methodology that aims to obtain a stable estimate of some characteristic defined by geographical regions or sociodemographic variables for each area or domain for which the sample size is not very large.
Along with the requirements of a wide range of practices, such as official statistics, public health, and business, recent progress in the research of small area estimation has been remarkable. 
Model-based small area estimation methods assume a class of mixed-effects models, which exploit the effect of `borrowing strength' from other areas.
Representative small area models are the Fay-Herriot model \citep{fay1979estimates} and the nested error regression model \citep{battese88}.
The former is an `area-level model' for aggregate data at the area-level, whereas the latter is a `unit-level model' for individual data.
For the comprehensive review of the small area estimation, see \cite{rao2015small} and \cite{sugasawa2020small}.

Model-based small area estimates are often required to be compatible with the estimates at a higher level of aggregation, which are usually direct estimates based on a survey sample.
For example, in the problem of estimating population totals, the model-based subnational estimates, when summed up, must be equal to the national estimates.
Such agreement is often demanded in official statistics not only for technical coherence but also for institutional and political reasons; for instance, an overall consistency with aggregate direct estimates may be politically necessary to demonstrate the utility of small area estimates to decision makers \citep{datta2011bayesian}.
An estimate at a higher level of aggregation is more reliable than small area estimates because the sample size is often large enough.
The method of estimating small area parameters using an estimate at a higher aggregation level as a benchmark is called the benchmarking method.
Since a benchmark at a higher aggregation level is reliable, the benchmarking method is not only used for practical reasons.
However, it is also known to provide some protection against model misspecification \citep{you2004, datta2011bayesian, steorts2013estimation}.

Various benchmarking methods have been proposed in a large body of literature.
In the point estimation theory, a natural approach is to construct the estimator based on a constrained risk minimization problem.
From a frequentist point of view, \cite{wang2008small} derived the benchmarked estimator as a unique constrained risk minimizer among a class of linear unbiased predictors.
\cite{bell2013benchmarking} extended this approach to problems with multiple benchmark constraints with a comprehensive discussion of internal and external benchmarking.
\cite{ghosh2019survey} reviews frequentist constrained estimation methods in finite samples, highlighting their relation to unconstrained approaches and unified optimization tools. 
Recently, \cite{chikamatsu2023benchmarked} derived the estimator that minimizes the constrained risk in the class of linear shrinkage estimators.

On the other hand, from the perspective of Bayesian decision theory, the point estimator is the minimizer of the posterior risk given the observed data, which is called the Bayes estimator.
When the squared loss is adopted as the loss function, the Bayes estimator becomes the posterior mean.
\cite{datta2011bayesian} constructed benchmarked estimators using the constrained Bayes method \citep{ghosh92cb} as a posterior risk minimization under the constraint of the benchmark.
\cite{kubokawa2014measuring} extended this approach to the class of natural exponential family models, and \cite{ghosh2015benchmarked} 
proposed the constrained Bayes estimator under the weighted K\"{u}llback-Leibler loss in the multiplicative area-level model.
\cite{ghosh2013two} developed a two-stage benchmarking method that satisfies constraints at both the area and unit levels in the context of the constrained Bayes method.
Recently, \cite{steorts2020smoothing} adopted a loss function that requires smoothness with respect to similarity across areas as well as benchmark constraints.

Some studies have tackled the benchmarking problem from a fully Bayesian approach, such as \cite{zhang2020fully} and \cite{okonek2022computationally}.
One of fully Bayesian benchmarking methods considers restricting the parameter space itself to satisfy the benchmark constraints. 
As a result, the posterior becomes degenerate, and sampling from the degenerate posterior is computationally challenging.
The constrained parameter space in benchmarking was also adopted by \cite{pfeffermann2006small} for the state-space model.
Another practical approach is to project the unconstrained posterior samples to the space where the benchmarking constraint is met. 
Such an approach would be more tractable as long as the computation of the benchmarked estimator is straightforward. 
However, it lacks theoretical foundations, and it might be difficult to interpret the confidence intervals based on such samples via an ad-hoc adjustment.

Our proposed fully Bayesian benchmarking method is built upon the information theoretic approach called entropic tilting \citep[e.g.,][]{jaynes1957information}.
The entropic tilting is a general method that gives the optimal distribution that minimizes the K\"{u}llback-Leibler divergence from the original posterior among a class of distributions whose expectations satisfy the given constraints. 
\cite{tallman2022entropic} and \cite{west2023perspectives} give recent reviews and examples using entropic tilting in Bayesian inference.
The proposed method does not constrain the parameter space but instead modifies the posterior distribution in such a way that the modified posterior expectation satisfies the benchmark constraints.

The proposed method provides the modified posterior distribution, which can provide point estimates as well as measures of uncertainty, such as credible intervals.
Conventional benchmarking methods have also evaluated the MSE of the point estimator and derived its asymptotically unbiased estimator \citep[e.g.][]{steorts2013estimation, kubokawa2014measuring}, under the assumption that the model is correctly specified. 
This is philosophically different from our uncertainty quantification based on the posterior distribution.
The distinction from frequentist MSE is that the posterior variance represents uncertainty conditional on the assumed model, while the posterior-based uncertainty quantification also relies on the modeling assumptions. 
It should also be noted that the proposed method is computationally feasible, which only requires a root-finding step for the parameters in tilting posterior and the computation is efficiently carried out by importance sampling.  
Hence, the computational complexity would be similar to that of the projected posterior approaches.
\cite{janicki2017} also adopted a method similar to ours to modify the posterior distribution, but they developed an approximation method that assumes that the original posterior is normally distributed.
On the other hand, we exactly evaluate the posterior distribution in hierarchical Bayesian models and thus produce more appropriate results than their method, especially when the original posterior is far from the normal distribution.
We consider a wider class of models, which is a general hierarchical model based on the natural exponential family proposed by \cite{ghosh2004small}, and propose a novel fully Bayesian benchmarking approach via entropic tilting of the posterior distribution.
Although there exists a variety of studies on the hierarchical models in the context of small area estimation, such as constructing confidence intervals \citep{ghosh2008empirical}, 
flexible latent distribution \citep{sugasawa2017empirical}, 
and extension to spatial data \citep{sugasawa2020spatial} to mention a few, 
only \cite{kubokawa2014measuring} considered the benchmarked estimation based on this model.

The rest of the paper is organized as follows.
Section~2 first introduces entropic tilting to the small area model based on the natural exponential family and derives the benchmarked posterior. 
Then, a general Monte Carlo method to obtain the benchmarked posterior is described.
Section~3 demonstrates that the benchmarked posteriors can be obtained in explicit forms for some representative small area models. 
Section 4 illustrates the numerical performance of the proposed method.
Finally, Section 5 provides some concluding remarks.

\section{Bayesian Benchmarking Estimation via Entropic Tilting}

\subsection{Small area models based on natural exponential family}\label{sec:model}

Let $y_i \ (i=1,\dots,m)$ be the response variable for  $i$th area and $x_i$ be the vector of associated covariates.
We consider the following general hierarchical model based on the natural exponential family:  
\begin{equation}\label{model}
\begin{split}
p(y_i|\theta_i)&= \exp \big[ \xi_i\{\theta_i y_i -\psi(\theta_i)\}+C_1(y_i, \xi_i)\big],\\
p(\theta_i|\bta)&=\exp \big[\nu\{m_i \theta_i-\psi\left(\theta_i\right)\}\big] C_2(\nu, m_i),
\end{split}
\end{equation}
where $\xi_i > 0$ is a known constant, and $C_1(\cdot,\cdot)$ and $C_2(\cdot,\cdot)$ are some functions specific to each distribution.
We consider the canonical link function  $m_i=\psi'(\bx_i^\top\bbe)$ with the vector of unknown regression coefficients $\bbe$.
Moreover, $\bta=(\bbe, \nu)$ is the vector of unknown parameters. 
Here, $E[y_i|\theta_i]=\psi'(\theta_i)\equiv \mu_i$, ${\rm Var}(y_i|\theta_i)=\psi''(\theta_i)/\xi_i$, $E[\psi'(\theta_i)|\bta]=m_i$. 
Since ${\rm Var}(y_i|\theta_i)>0$, $\mu_i$ is strictly increasing in $\theta_i$.
The model \eqref{model} is known to include representative small area models such as Fay-Herriot model \citep{fay1979estimates} and Poisson-gamma model \citep{clayton1987empirical}.

For hierarchical Bayesian (HB) approaches, let $\pi(\bta)$ be a prior distribution of $\bta$ and the joint posterior of $\theta_i$ and $\bta$ is given by 
\begin{equation}\label{eq:pos}
p(\theta_1,\ldots,\theta_m,\bta|y)\propto \pi(\bta) \left[ \prod_{i=1}^mp(\theta_i|\bta) p(y_i|\theta_i) \right],
\end{equation}
where $y$ denotes the collection of $y_i$. 
The conditional posterior of $\theta_i$ is mutually independent over $i=1,\ldots,m$, and it is given by
\begin{equation}\label{theta_conditional}
p(\theta_i|\bta, y_i)\propto \exp \big\{ \theta_i(\xi_i y_i+\nu m_i)-(\xi_i+\nu)\psi(\theta_i)\big\}, \quad i=1,\ldots,m.
\end{equation}
Note that the above conditional posterior belongs to the same family as the conditional prior distribution of $\theta_i$ given $\bta$. 
Then, the conditional posterior mean of $\mu_i = \psi'(\theta_i)$ given $\bta$ is 
$$
\widetilde{\mu}_i(\bta)\equiv E[\psi'(\theta_i)|y_i, \bta]=\frac{\xi_i y_i+\nu m_i}{\xi_i+\nu}.
$$
Hence, the marginal posterior mean of $\mu_i$ is $\widehat{\mu}_i=E_{\eta}[\widetilde{\mu}_i(\bta)]$, where $E_{\eta}$ denotes the expectation with respect to the marginal posterior $\pi(\bta|y)$ of $\bta$.

Alternatively, one can obtain the point estimate $\widehat{\bta}_{\rm ML}$ of $\bta$ by maximizing the marginal log-likelihood
$$
\widehat{\bta}_{\rm ML}={\rm argmax}_{\bta}\sum_{i=1}^m \log\int p(\theta_i|\bta) p(y_i|\theta_i) d\theta_i,
$$
and define ``empirical" posterior of $\theta_i$ as (\ref{theta_conditional}) with $\bta=\widehat{\bta}_{\rm ML}$.
This can be regarded as replacing the marginal posterior of $\bta$ with a Dirac measure on $\bta=\widehat{\bta}_{\rm ML}$, which we call empirical Bayes (EB) approaches.

\subsection{Benchmarking via entropic tilting}
Here, we consider the hierarchical Bayes estimation under constraints via entropic tilting (ET). 
Before introducing the proposed Bayesian benchmarking method, we first describe the ET approach briefly.

Suppose for now that we would like to introduce constraints to the density function $p(\bth)$ where $\bth=(\theta_1,\dots,\theta_m)$. 
The ET framework seeks a probability density under which the constraint in the form $E_g[s_j(\bth)]=0$ is met for $j=1,\dots,J$, where $s_j(\bth)$ is a known function of $\bth$. 
Then ET chooses the tilted distribution $g(\bth)$ as the closest density to the baseline $p(\bth)$ subject to the constraints in the sense of K\"{u}llback-Leibler (KL) divergence: 
$$
{\rm KL}_{g|p}=\int \log\frac{g(\bth)}{p(\bth)} g(\bth)d\bth. 
$$
Provided a solution exists, it is known to be in the unique form given by $g(\bth)\propto \exp(\bga^\top \bs(\bth))p(\bth)$, where $\bs(\bth)=(s_1(\bth),\dots,s_J(\bth))^\top$ and $\bga$ is the parameter vector that ensures $g(\bth)$ satisfies the constraints. 
Examples include standard constraint for the mean in the form $\sum_{i=1}^m w_iE[\psi'(\theta_i)\mid y]=C$ for some fixed weights $w_i$ and constant $C$, corresponding to $s_1(\bth)=\sum_{i=1}^m w_iE[\psi'(\theta_i)\mid y]-C$ and $J=1$. 
According to general results on information projections \citep{csiszar1975divergence}, a solution exists whenever the imposed constraint lies inside the convex hull of the values that the baseline posterior distribution can take.
This means that if the target constraint is located within the range of what the baseline distribution can plausibly produce, then the tilted distribution is guaranteed to exist.
See \cite{jaynes1957information} for more details and \cite{tallman2022entropic} for the use of entropic tilting in Bayesian forecasting.

Under the mean constraint given above, the tilted posterior distribution of $(\theta_1,\ldots,\theta_m)$ is given by 
\begin{equation}\label{ET-pos}
\begin{split}
&g(\theta_1,\ldots,\theta_m|y)\\
& \ \ \propto \exp \Big\{ \gamma \Big(\sum_{i=1}^m w_i \psi'(\theta_i)-C\Big) \Big\} \int \Big[ \prod_{i=1}^m
\exp \big\{ \theta_i(\xi_i y_i+\nu m_i)-(\xi_i+\nu)\psi(\theta_i)\big\} \Big] \pi(\bta|y)d\bta,
\end{split}
\end{equation}
where $\gamma$ is the solution of the equation $\sum_{i=1}^m w_iE_g[\psi'(\theta_i)]=C$ and $E_g$ denotes the expectation with respect to the tilted posterior (\ref{ET-pos}).
In general, the above distribution does not admit analytical forms.
However, under EB approaches using a Dirac measure on $\bta=\widehat{\bta}_{\rm ML}$ for $\pi(\bta|y)$ in (\ref{ET-pos}), some models lead to analytical posterior distributions.  
Particularly, we will demonstrate that this is the case for the Fay-Herriot and Poisson-gamma models in Section~\ref{sec:illust}.
For those models, $g_{\eta}(\theta_i)$ is in the same parametric family as the conditional prior distribution of $\theta_i$ given $\bta$, and the EB tilted posterior is in a relatively simple form. 
More specifically, $\psi'(\theta_i) = \theta_i$ for the Fay-Herriot model and $\psi'(\theta_i) = \psi(\theta_i)$ for the Poisson-gamma model.

\subsection{A Monte Carlo algorithm for the tilted posterior}
As noted above, a closed form for the tilted posterior \eqref{ET-pos} is not usually available, particularly for HB approaches.  
Here, a general Monte Carlo algorithm for solving for the tilting parameter $\gamma$ and simulating from the tilted posterior \eqref{ET-pos} is described.
We consider a general case with multiple constraints, $E_g[s_j(\bth)]=0$ for $j=1,\ldots,J$, and where the expectation is taken with respect to the tilted posterior with given $\bga$, and $\bga$ is defined as the solution of the multiple equations. 
To generate random samples of $\theta_1,\dots,\theta_m$ from \eqref{ET-pos} with fixed $\gamma$, we may use importance sampling.
Based on the posterior samples, $\{\bth^{(1)},\dots,\bth^{(S)}\}$ from the original posterior (\ref{eq:pos}), we define the unnormalized importance weight as $\omega^{(s)}(\bga)=\exp\{\bga^\top \bs(\bth^{(s)})\}$ for $s=1,\ldots,S$. 
The posterior samples from the tilted posterior (\ref{ET-pos}) can be generated by a multinomial distribution on $\{\bth^{(1)},\dots,\bth^{(S)}\}$ with the vector of probabilities proportional to $\left\{\omega^{(1)}(\bga),\dots,\omega^{(S)}(\bga)\right\}$. 
Hence, the equation for $\bga$ can be approximated as
\begin{equation}\label{snis}
\sum_{s=1}^S\omega^{(s)}(\bga) s_j(\bth^{(s)})=0,
\end{equation}
for $j=1,\ldots,J$.
This expression does not require additional intensive computation, such as a Markov chain Monte Carlo method, so that the calculation of the optimal $\bga$ is tractable. 
The detailed steps for the computation of $\bga$ is summarized as follows: 

\begin{algo}
Based on the posterior samples $\{\bth^{(1)},\dots,\bth^{(S)}\}$ from the original posterior (\ref{eq:pos}), the optimal $\bga$ can be obtained as follows: 
\begin{enumerate}
\item 
Set $\bga^{(0)}=\zero$ and $t=0$. 

\item 
Given $\bga^{(t)}$, update $\bga$ by the following Newton-Raphson algorithm for (\ref{snis}): 
$$
\bga^{(t+1)}  \ \ \leftarrow \ \ 
\bga^{(t)}- \left\{\sum_{s=1}^S\omega^{(s)}(\bga^{(t)})\bs(\bth^{(s)})\bs(\bth^{(s)})^\top \right\}^{-1} 
\sum_{s=1}^S\omega^{(s)}(\bga^{(t)})\bs(\bth^{(s)}).
$$

\item 
If $\|\bga^{(t+1)}-\bga^{(t)}\|<\delta$ for some small $\delta>0$ (e.g. $\delta=10^{-5}$), the algorithm is terminated; otherwise, set $t=t+1$ and go to step 2. 
\end{enumerate}
\end{algo}

Let us denote the solution by $\tilde{\bga}$. 
The unweighted sample from the tilted posterior can be obtained by sampling from $\big\{\bth^{(1)},\dots,\bth^{(S)}\big\}$ with the associated weights  $\big\{\tilde{\omega}^{(1)},\dots,\tilde{\omega}^{(S)}\big\}$ where $\tilde{\omega}^{(s)}=\omega^{(s)}(\tilde{\bga})/\sum_{s'=1}^S\omega^{(s')}(\tilde{\bga})$.

\subsection{Approximation error of the tilted posterior}
Here, we discuss the approximation error of the tilted posterior.
We consider $J$ constraints of the form, $s_j(\bth)=\sum_{i=1}^m w_i h_j(\theta_i)$ \ ($j=1,\ldots,J$) for some function $h_j(\cdot)$, as typically appeared in the benchmarking constraints.
For example, the mean constraint is written as $h_j(\theta_i)=\psi'(\theta_i)- C_j$ for some constant $C_j$. 
In this setting, the $J$ constraints can be expressed as $\bs(\bth)=\sum_{i=1}^m w_i \bh(\theta_i)=\zero$, where $\bh(\theta_i)=(h_1(\theta_i),\ldots,h_J(\theta_i))^\top$. 
Then, the tilted posterior density $g(\bth|\by)$ is given by 
$$
g(\bth|\by)=V_m^{-1} \int \prod_{i=1}^m \exp\{ w_i \bga^\top \bh(\th_i) \} f(\theta_i|\bta,\by)\pi(\bta|\by)d\bta, 
$$
where 
$$
V_m=\int \left[\prod_{i=1}^m \int\exp\{w_i \bga^\top \bh(\th_i) \}f(\theta_i|\bta,\by)d\theta_i\right]\pi(\bta|\by)d\bta.
$$
and $\gamma$ is the solution of $\sum_{i=1}^m w_i\bh(\theta_i)=\zero$.
Then, we have the following results regarding approximation errors of the tilted posterior density, where the proof is deferred to the Appendix.

\begin{thm}\label{thm:approx}
Under $w_i=O(m^{-1})$, it holds that ${\rm KL}(g(\bth|\by)|f(\bth|\by))=O(m^{-1})$.  
\end{thm}

The assumption $w_i=O(m^{-1})$ is typically satisfied in the benchmarking problems as $w_i$ is typically a standardized weight such that $\sum_{i=1}^m w_i=1$. 
Regarding $C$, we indeed treat it as a known fixed value in the benchmarking constraint. 
The result in Theorem~1 is in parallel with the classical finding that the difference of mean squared errors between the empirical Bayes estimator and its benchmarked version is of order $O(m^{-1})$ as $m\to\infty$ \citep[e.g.][]{steorts2013estimation, kubokawa2014measuring}. 
Our Theorem 1 can be regarded as a distributional analogue of this result, extending the $O(m^{-1})$ difference from the level of point estimators to the level of posterior distributions.

\section{Illustrative Models and Benchmarking Constraints}
\label{sec:illust}

\subsection{General Fay-Herriot model with mean constraint}\label{sec:ex-FH}
Fay-Herriot model \citep{fay1979estimates} is recognized as the gold standard model in small area estimation. 
The general Fay-Herriot model is expressed as 
\begin{equation}\label{FH}
y_i|\theta_i \sim N(\theta_i, D_i), \quad \theta_i\sim N(\bx_i^\top\bbe, Au_i),  \quad u_i\sim \pi_\lambda, \ \quad i=1,\ldots,m,
\end{equation}
where $D_i$ is fixed and $\bta=(\bbe^\top, A)^\top$ is the vector of unknown parameters.
Here $u_i$ is the local parameter that controls the degree of shrinkage, and follows a user-specified distribution $\pi_\lambda$, which may depend on some unknown parameter $\lambda$. 
By integrating $u_i$ out, the marginal model of $\theta_i$ is a scale mixture of normals. 
See \cite{tang2018modeling} for such a general class of priors for $\theta_i$. 
A representative example includes $u_i\sim {\rm Exp}(\lambda^2/2)$, leading to the Laplace prior \citep{park2008bayesian} and $u_i\sim C^{+}(0, 1)$ leading to the horseshoe prior \citep{carvalho2010horseshoe} for $\theta_i$.   

Given $u_i$ and $\bta$, the posterior distribution of $\theta_i$ is $N(\tht_i(u_i,\bta), \sit_i^2(u_i,\bta))$, where 
\begin{equation}\label{FH-pos}
\tht_i(u_i,\bta)=y_i- \frac{D_i}{u_iA+D_i}(y_i-\bx_i^\top\bbe), \ \ \ \ \sit_i^2(u_i,\bta)=\frac{u_iAD_i}{u_iA+D_i}.
\end{equation}
We then define $\pi(\bu,\bta|\by)$ for $\bu=(u_1,\ldots,u_m)$ and $\by=(y_1,\ldots,y_m)$ as the posterior distribution of $(\bu,\bta)$ given by 
$$
\pi(\bu,\bta|\by)=\frac{\int \prod_{i=1}^m \phi(y_i;\theta_i, D_i)\phi(\theta_i; \bx_i^\top\bbe, u_iA)\pi(u_i;\lambda)\pi(\bta) d\bth }{\iiint \prod_{i=1}^m \phi(y_i;\theta_i, D_i)\phi(\theta_i; \bx_i^\top\bbe, u_iA)\pi(u_i;\lambda)\pi(\bta) d\bth d\bu d\bta},
$$
where $\pi(\bta)$ is the prior distribution for $\bta$. 
To generate a sample from the marginal posterior $\pi(\bu,\bta|y)$, we typically rely on a Markov Chain Monte Carlo algorithm, where the details are given in the Appendix. 
Based on the posterior $\pi(\bu,\bta|\by)$, the marginal posterior of $\bth=(\theta_1,\ldots,\theta_m)$ can be written as 
$$
f(\bth|\by)=\int \prod_{i=1}^m \phi(\theta_i;\tht_i(u_i,\bta), \sit_i^2(u_i,\bta)) \pi(\bu,\bta|\by)d\bu d\bta.
$$
The posterior means of $\theta_i$ under the above posterior do not necessarily satisfy a constraint such as $\sum_{i=1}^m w_i\thh_i=C$, where $C$ is a known fixed value and $\thh_i=E_f[\theta_i]$ with the expectation taken with respect to $f(\bth|\by)$. 
We assume that $w_i=O(1/m)$ and $C=O(1)$.

Our goal is to find the closest posterior distribution (in terms of KL divergence) to the unconstrained posterior $f(\bth|\by)$ under some moment conditions. 
We first consider a typical situation of benchmarking with the mean constraint, $\sum_{i=1}^m w_i\theta_i=C$ for some fixed value $C$. 
According to the general results of the entropic tilting, the optimal form of the constrained posterior distribution is given by 
$$
g(\bth|\by)\propto \exp\Big\{\gamma \Big(\sum_{i=1}^m w_i\theta_i-C\Big)\Big\}f(\bth|\by),
$$
where $\gamma>0$ is the tilting parameter to ensure that $\sum_{i=1}^m w_iE_g[\theta_i]=C$ with the expectation taken with respect to the tilted distribution $g(\bth)$.
In HB settings, we can use Algorithm~1 to obtain $\gamma$ and random samples from the tilted posterior.

Next, we consider an EB approach.
Let $\widehat{\bu}$ and $\widehat{\bta}$ be posterior means of $\bu$ and $\bta$ based on the posterior samples. 
Then, an empirical posterior of $\theta_i$ can be defined as $N(\tht_i(\widehat{u}_i,\widehat{\bta}), \sit_i^2(\widehat{u}_i,\widehat{\bta}))$.
Note that the EB tilted posterior can be expressed as 
\begin{align}
g(\bth|\by)
&\propto \exp\Big\{\gamma \Big(\sum_{i=1}^m w_i\theta_i-C\Big)\Big\}\prod_{i=1}^m \phi(\theta_i;\tht_i(\widehat{u}_i,\widehat{\bta}), \sit_i^2(\widehat{u}_i,\widehat{\bta}))\notag \\
&\propto\ 
\prod_{i=1}^m\phi\Big(\theta_i;\tht_i(\widehat{u}_i,\widehat{\bta})+\gamma w_i\sit_i^2(\widehat{u}_i,\widehat{\bta}), \sit_i^2(\widehat{u}_i,\widehat{\bta})\Big).
\label{pos-approx}
\end{align}
The tilting parameter $\gamma$ is determined by the constraint $\sum_{i=1}^m w_iE_g[\theta_i]=C$, which gives a closed-form solution of $\gamma$ as 
\begin{equation}\label{gamma_hf}
\gamma=\frac{C-\sum_{i=1}^m w_i\tht_i(\widehat{u}_i,\widehat{\bta})}{\sum_{i=1}^mw_i^2\sit_i^2(\widehat{u}_i,\widehat{\bta})}.
\end{equation}
Then, the resulting EB benchmarked estimator (posterior mean of the tilted posterior) is given by
\begin{equation}\label{bench_hf}
\thh_i^{(B)}=\tht_i(\widehat{u}_i,\widehat{\bta})+\frac{w_i\sit_i^2(\widehat{u}_i,\widehat{\bta})}{\sum_{i'=1}^m w_{i'}^2\sit_{i'}^2(\widehat{u}_{i'},\widehat{\bta})}\left(C-\sum_{i'=1}^m w_{i'}\tht_{i'}(\widehat{u}_{i'},\widehat{\bta})\right),
\end{equation}
which satisfies $\sum_{i=1}^m w_i\thh_i^{(B)}=C$.
A similar form can be obtained through the constrained weighted quadratic loss with weight being $\sit_{i}^2(\theta)$ \citep[e.g.][]{datta2011bayesian}.

\subsection{Fay-Herriot model with both first- and second-moment constraints}
Sometimes, one is interested in benchmarked estimation under both first- and second-moment constraints.
For simplicity, we here assume that $u_i=1$. 
We consider an additional constraint: $\sum_{i=1}^m w_iE_f[\{\theta_i-\tht_i(\bta)\}^2]=H$ for some constant $H$.
Using entropic tilting, the optimal approximate posterior under two constraints is given by 
\begin{align*}
g(\bth|\by)
&\propto \int \left[ \prod_{i=1}^m \exp\left\{\gamma_1 w_i\theta_i -\frac12\gamma_2w_i(\th_i-\tht_i(\bta))^2 \right\}\phi(\theta_i;\tht_i(\bta), \sit_i^2(\bta))\right] \pi(\bta|y)d\bta.
\end{align*}
Then, the optimal $\gamma=(\gamma_1,\gamma_2)$ is determined by Algorithm~1 for HB approaches.

Regarding the EB approach, the EB tilted posterior is expressed as 
\begin{equation} \label{pos-approx2}
\begin{split}
&\prod_{i=1}^m \exp\left\{\gamma_1 w_i\theta_i -\frac12\gamma_2w_i(\th_i-\tht_i(\widehat{\bta}))^2 \right\}\phi(\theta_i;\tht_i(\widehat{\bta}), \sit_i^2(\widehat{\bta}))\\
&\propto
 \prod_{i=1}^m\phi\left(\theta_i;  \frac{\tht_i(\widehat{\bta})+w_i\sit_i^2(\widehat{\bta})(\gamma_1+\gamma_2\tht_i(\widehat{\bta}))}{1+\gamma_2w_i\sit_i^2(\widehat{\bta})},   \frac{\sit_i^2(\widehat{\bta})}{1+\gamma_2w_i\sit_i^2(\widehat{\bta})}   \right).
\end{split}
\end{equation}
and the tilting parameters, $\gamma_1$ and $\gamma_2$, are determined by the following two constraints:
\begin{align*}
E_g\left[\sum_{i=1}^m w_i\theta_i\right]=\sum_{i=1}^m \frac{w_i\tht_i(\widehat{\bta})+w_i^2\sit_i^2(\widehat{\bta})(\gamma_1+\gamma_2\tht_i(\widehat{\bta}))}{1+\gamma_2w_i\sit_i^2(\widehat{\bta})}=C
\end{align*}
and 
\begin{align*}
E_g\left[\sum_{i=1}^m w_i(\theta_i-\tht_i(\widehat{\bta}))^2\right]
&=\sum_{i=1}^m w_i{\rm Var}_g(\theta_i) + \sum_{i=1}^m w_i (E_g[\theta_i]-\tht_i(\widehat{\bta}))^2\\
&=\sum_{i=1}^m  \frac{w_i\sit_i^2(\widehat{\bta})}{1+\gamma_2w_i\sit_i^2(\widehat{\bta})}
+\sum_{i=1}^m w_i\left(\frac{\gamma_1w_i\sit_i^2(\widehat{\bta})}{1+\gamma_2w_i\sit_i^2(\widehat{\bta})} \right)^2 = H.
\end{align*}
By numerically solving the above two equations for $\gamma_1$ and $\gamma_2$, we can obtain the benchmarked estimator as well as the approximate EB posterior distribution for uncertainty quantification.

\subsection{Poisson-gamma model under mean constraint}
The Poisson-gamma model is expressed as 
\begin{equation}\label{PG}
y_i = z_i / n_i, \ \ \ z_i \sim {\rm Po}(n_i\la_i), \ \ \ \la_i\sim {\rm Ga}(\nu m_i, \nu), \ \ \ \ i=1,\ldots,m, 
\end{equation}
where $m_i=\exp(\bx_i^\top\bbe)$, $\mathrm{Ga}(a,b)$ denotes the gamma distribution with the shape parameter $a$ and rate parameter $b$, $\bta=(\bbe^\top, \nu)^\top$ is the vector of unknown parameters and $n_i$ is the known constant. 
Then, this model corresponds to \eqref{model} with $\theta_i = \log(\la_i)$ and $\psi(\theta_i) = \exp(\theta_i)$.
Given $\bta$, the conditional posterior distribution of $\la_i$ is ${\rm Ga}(n_iy_i+\nu m_i, n_i+\nu)$, and the conditional posterior mean of $\la_i$ is $E[\la_i|\bta]=(n_i+\nu)^{-1}(n_iy_i+\nu m_i)$. 
We consider the mean constraint given by $\sum_{i=1}^m w_i E_f[\la_i]=C$, where the expectation is with respect to the joint posterior $f$ of $\bla=(\la_1,\dots,\la_m)$.
Then, the optimal approximation of the posterior under the mean constraint can be expressed as 
$$
g(\bla)
\propto \exp\Big\{\gamma \Big(C-\sum_{i=1}^m w_i\la_i\Big)\Big\}\int \Big[ \prod_{i=1}^m p_{\rm Ga}(\la_i;n_iy_i+\nu m_i, n_i+\nu) \Big] \pi(\bta|\by)d\bta.
$$
For the HB approach, the tilting parameter $\gamma$ in the above posterior can be computed by Algorithm~1.

For the EB approach, we let $\widehat{\nu}$ and $\widehat{m}_i$ be posterior means of $\nu$ and $m_i$, respectively. 
Then, the EB tilted posterior is 
\begin{align*}
g(\bla) 
&= \exp\Big\{\gamma \Big(C-\sum_{i=1}^m w_i\la_i\Big)\Big\} \prod_{i=1}^m p_{\rm Ga}(\la_i;n_iy_i+\widehat{\nu} \widehat{m}_i, n_i+\widehat{\nu})\\
&=\prod_{i=1}^mp_{\rm Ga}(\la_i;n_iy_i+\widehat{\nu} \widehat{m}_i, n_i+\widehat{\nu}+\gamma w_i),
\end{align*}
with the tilting parameter $\gamma$ determined by the solution of the following equation: 
\begin{equation}\label{gamma_pg}
\sum_{i=1}^m w_i E_g[\psi'(\theta_i)]= \sum_{i=1}^m \frac{w_i(n_iy_i+\nu \exp(\bx_i^\top \bbe))}{n_i+\nu+\gamma w_i}=C.
\end{equation}
Since the left-hand side of (\ref{gamma_pg}) is a decreasing function of $\gamma$ and its range is $(0, \infty)$, the equation (\ref{gamma_pg}) has a unique solution. 
By solving the above equation via some iterative algorithms (e.g. Newton-Raphson algorithm), one can obtain the solution $\tilde{\gamma}$ and gives the EB tilted posterior.

\section{Numerical studies}
\subsection{Simulation study 1: Fay-Herriot model}

We first demonstrate the numerical performance of the proposed entropic tilting approach under the Fay-Herriot model. 
In particular, we focus on showing that the entropic tilting approach provides benchmarked point estimates as well as reasonable uncertainty quantification under general prior specifications for random effects, as provided in Section~\ref{sec:ex-FH}.

We set $m=30$ and generate two covariates, $x_{i1}$ and $x_{i2}$, from $U(-2,2)$ and ${\rm Ber}(0.5)$, respectively, for $i=1,\ldots,m$. 
For sampling variance $D_i$ and area-specific sample size $n_i$, we divided $m$ areas into five groups with equal numbers of areas. 
The areas within the same group share the same sampling variance and area-specific sample size.
The group patterns are $(0.3, 0.7, 1, 1.5, 2.0)$ for $D_i$ and $(30, 20, 15, 10, 5)$ for $n_i$. 
We then generate $\theta_i$ from the following two data-generating process: 
\begin{align*}
{\rm (A)}\ \ \theta_i=-2+x_{i1}+x_{i2}+\sqrt{A} v_i, \ \ \ \ \ \ \ 
{\rm (B)}\ \ \theta_i=-2 + x_{i1}^2 + x_{i2}x_{i1}^2 +\sqrt{A} v_i
\end{align*}
and generate $y_i$ from $y_i \mid \theta_i \sim N(\theta_i, D_i)$.
We set $A=0.5$.
Regarding the distribution of $v_i$, we consider the following three scenarios: 
$$
{\rm (I)} \  v_i\sim N(0, 1), \ \ \ \
{\rm (II)} \  v_i\sim 0.7N(0,1)+0.3\delta_0, \ \ \ \
{\rm (III)} \  v_i\sim \sqrt{(\nu-2)/\nu}t_\nu,
$$
where $\delta_0$ is the Dirac distribution on the origin and $t_\nu$ denotes the $t$-distribution with $\nu$ degrees of freedom with $\nu=2.5$.
Note that there are some areas without random effects in scenario (II), which is known as uncertain random effects \citep{datta2015small}, and the data may contain outlying values in scenario (III). 
Since we only use $(x_{i1}, x_{i2})$ in the model, the FH model is correctly specified in scenario (A), while not under scenario (B).

For estimating $\theta_i$, we fitted the Fay-Herriot model
$$
y_i|\theta_i\sim N(\theta_i, D_i), \ \ \ \theta_i=\beta_0+\beta_1 x_{i1} + \beta_2x_{i2} + v_i, 
$$
and adopted the normal (N) and horseshoe \citep[HS;][]{carvalho2010horseshoe} distributions for the random effect $v_i$. 
It is noted that the both cases are misspecified in (II) and (III).  
We generated 8000 posterior draws after discarding the first 2000 samples as the burn-in period, and computed the hierarchical Bayes (HB) estimator of $\theta_i$ given by the posterior mean. 
Then, we consider an external benchmarking constraint given by $\sum_{i=1}^m w_i\theta_i=C$ with $C=m^{-1}\sum_{i=1}^m \theta_i^{\ast}$, where $w_i=n_i/\sum_{i'=1}^m n_{i'}$ and $\theta_i^{\ast}$ is the true value. 
To meet the constraint, we applied the proposed hierarchical Bayes and empirical Bayes entropic tilting, denoted by HET and EET, respectively, to obtain the benchmarked posterior. 
For comparison, we employed the Bayesian benchmarking method based on minimum discrimination information (MDI) with a normal approximation of the posterior, as proposed by \cite{janicki2017}.
Furthermore, we considered the constraint Bayes estimator \citep{datta2011bayesian} given by $\widehat{\theta}_i^\mathrm{CB} = \widehat{\theta}_i + w_i(C - \sum_{i=1}^m w_i\widehat{\theta}_i)/s \phi_i$ where $s=\sum_{i=1}^mw_i^2\phi_i$, with $\phi_i={\rm Var}(\theta_i|\by)^{-1}$ (WL1) and $\phi_i=D_i^{-1}$ (WL2). 
Based on WL2, we also adopted the posterior projection (PP) approach that applies the benchmarking constraint to each posterior sample.  
Since the four Bayesian methods (HET, EET, MDI, and PP) provide posterior samples satisfying the benchmarking constraint, we compute not only point estimates (as posterior means) but also $95\%$ credible intervals of $\theta_i$.

To quantify the performance of point estimation, we computed the mean squared errors (MSE), defined by $m^{-1}\sum_{i=1}^m (\widehat{\theta}_i - \theta_i)^2$ for an estimate $\widehat{\theta}_i$.  
Moreover, to evaluate uncertainty quantification, we evaluated the coverage probability (CP) and average length (AL). 
The results are given in Tables~\ref{tab:sim-mse} and \ref{tab:sim-CI}.

Table~\ref{tab:sim-mse} shows that all methods resulted in higher MSE under the HS prior and in the data-generating process (B), where the fitted covariate structure is misspecified. 
It is seen that the MSE for the proposed HET achieved the lowest or tied-for-lowest in nearly all data-generating processes and prior settings for $\theta_i$. 
While the performance of WL2 is generally comparable to HET, the MSE for HET is marginally smaller in most cases. 
A common pattern is observed for PP and WL1, where the increase in MSE under the HS prior is larger than that for HET and WL2. 
By contrast, the performance of EET, in contrast, is similar to HB, which resulted in the largest MSE in all cases, with only marginal improvements. 

From Table~\ref{tab:sim-CI}, in Scenario (A), all methods resulted in CP close to 97\%, often exceeding the nominal 95\%. 
In Scenario (B), CP under the normal prior, CP is close to the nominal 95\% for all methods. 
On the other hand, under the HS prior, CP drops slightly to about 92\% across methods, indicating undercoverage when the covariate structure is misspecified. 

Turning to AL in Table~\ref{tab:sim-CI}, PP produced the shortest credible intervals, followed by MID and EET, both of which resulted in much narrower credible intervals than HET or HB under the HS prior. 
Notably, AL for HB and HET are essentially identical in most cases. 
This finding demonstrates that HET enforces the benchmark constraint through modifying the original HB posterior, thereby inheriting much of the uncertainty structure of the HB posterior. 
This stands in sharp contrast to PP, which introduces the constraint in an ad-hoc manner and hence distorts the quantification of uncertainty. 

The result that the proposed method based on ET performs reasonably well under various prior distributions suggests its broad applicability.

\begin{table}[tb]
\caption{Mean squared errors (MSE) of the hierarchical Bayes (HB) estimator of $\theta_i$ as well as the benchmarked estimator by hierarchical Bayesian entropic tilting (HET), empirical Bayesian entropic tilting (EET), approximated minimum discrimination information (MDI), posterior projection (PP) and two types of weighted loss functions (WL1 and WL2), under normal (N) and horseshoe (HS) priors for $\theta_i$, for the Fay-Herriot model. 
The results are based on 2000 Monte Carlo replications.  
\label{tab:sim-mse}
}
\begin{center}
{
\begin{tabular}{ccccccccccccccc}
\hline
 \multicolumn{2}{c}{DGP}  & Prior &  & HB & HET & EET & MDI & PP & WL1 & WL2 \\
\hline
 & (I) & N &  & 0.308 & 0.293 & 0.307 & 0.294 & 0.295 & 0.295 & 0.295 \\
 &  & HS &  & 0.364 & 0.349 & 0.363 & 0.350 & 0.355 & 0.355 & 0.350 \\
(A) & (II) & N &  & 0.214 & 0.198 & 0.213 & 0.198 & 0.199 & 0.199 & 0.199 \\
 &  & HS &  & 0.273 & 0.256 & 0.271 & 0.257 & 0.262 & 0.260 & 0.258 \\
 & (III) & N &  & 0.277 & 0.262 & 0.276 & 0.262 & 0.264 & 0.263 & 0.263 \\
 &  & HS &  & 0.317 & 0.302 & 0.315 & 0.302 & 0.307 & 0.307 & 0.303 \\
\hline
 & (I) & N &  & 0.791 & 0.774 & 0.789 & 0.774 & 0.775 & 0.775 & 0.774 \\
 &  & HS &  & 0.905 & 0.889 & 0.903 & 0.890 & 0.896 & 0.899 & 0.890 \\
(B) & (II) & N &  & 0.786 & 0.765 & 0.784 & 0.765 & 0.765 & 0.767 & 0.765 \\
 &  & HS &  & 0.906 & 0.885 & 0.904 & 0.886 & 0.890 & 0.898 & 0.886 \\
 & (III) & N &  & 0.798 & 0.781 & 0.797 & 0.781 & 0.781 & 0.782 & 0.780 \\
 &  & HS &  & 0.917 & 0.899 & 0.916 & 0.900 & 0.906 & 0.912 & 0.900 \\
\hline
\end{tabular}
}
\end{center}
\end{table}

\begin{table}[tb]
\caption{
Coverage probability (CP) and average length (AL) of $95\%$ credible intervals of $\theta_i$, based on the unconstrained hierarchical Bayes (HB) posterior as well as the benchmarked posterior by hierarchical Bayesian entropic tilting (HET), empirical Bayesian entropic tilting (EET), approximated minimum discrimination information (MDI) and posterior projection (PP), under normal (N) and horseshoe (HS) priors for $\theta_i$, for the Fay-Herriot model. 
The results are based on 2000 Monte Carlo replications.   
\label{tab:sim-CI}
}
\begin{center}
{
\begin{tabular}{ccccccccccccccc}
\hline
&&&& \multicolumn{5}{c}{CP (\%)} && \multicolumn{5}{c}{AL}\\
\multicolumn{2}{c}{DGP}   & Prior  &  & HB & HET & EET & MDI & PP &  & HB & HET & EET & MDI & PP \\
 \hline
 & (I) & N &  & 97.2 & 97.6 & 97.7 & 97.4 & 96.8 &  & 2.49 & 2.49 & 2.48 & 2.48 & 2.40 \\
 &  & HS &  & 96.7 & 97.0 & 97.7 & 97.4 & 96.0 &  & 2.72 & 2.72 & 2.66 & 2.66 & 2.62 \\
(A) & (II) & N &  & 98.5 & 98.8 & 98.9 & 98.7 & 98.4 &  & 2.38 & 2.38 & 2.36 & 2.36 & 2.28 \\
 &  & HS &  & 98.2 & 98.6 & 99.2 & 98.9 & 98.0 &  & 2.64 & 2.64 & 2.57 & 2.57 & 2.54 \\
 & (III) & N &  & 97.7 & 98.0 & 98.1 & 97.9 & 97.4 &  & 2.45 & 2.45 & 2.44 & 2.44 & 2.36 \\
 &  & HS &  & 97.8 & 98.1 & 98.6 & 98.4 & 97.2 &  & 2.68 & 2.67 & 2.61 & 2.61 & 2.58 \\
 \hline
 & (I) & N &  & 94.8 & 94.9 & 94.9 & 94.8 & 94.3 &  & 3.37 & 3.37 & 3.37 & 3.37 & 3.29 \\
 &  & HS &  & 92.6 & 92.8 & 92.0 & 91.7 & 91.8 &  & 3.36 & 3.36 & 3.35 & 3.35 & 3.27 \\
(B) & (II) & N &  & 94.7 & 94.9 & 94.9 & 94.7 & 94.2 &  & 3.35 & 3.35 & 3.35 & 3.35 & 3.28 \\
 &  & HS &  & 92.3 & 92.7 & 92.1 & 91.4 & 91.7 &  & 3.35 & 3.35 & 3.34 & 3.34 & 3.26 \\
 & (III) & N &  & 94.6 & 94.9 & 94.8 & 94.6 & 94.2 &  & 3.37 & 3.36 & 3.36 & 3.36 & 3.29 \\
 &  & HS &  & 92.3 & 92.7 & 91.9 & 91.4 & 91.6 &  & 3.36 & 3.35 & 3.35 & 3.35 & 3.26 \\
\hline
\end{tabular}
}
\end{center}
\end{table}

\subsection{Simulation study 2: Poisson-gamma model}\label{sec:PG-sim}
We next consider experiments with the Poisson-gamma model. 
We set $m=30$ and generated two covariates, $x_{i1}$ and $x_{i2}$, in the same way as in Section~\ref{sec:ex-FH}.
Then, the synthetic data is generated from $z_i|\lambda_i\sim {\rm Po}(\lambda_i n_i)$ with three scenarios of the true Poisson intensity given by 
\begin{align*}
&{\rm (I)} \  \lambda_i\sim {\rm Ga}(\nu m_i, \nu), \ \ \ \ \ 
{\rm (II)} \  \lambda_i\sim 0.8{\rm Ga}(\nu m_i, \nu)+0.2{\rm Ga}(3\nu m_i, \nu),  \\
&{\rm (III)} \  \lambda_i\sim {\rm LN}(\log m_i, \nu/20),
\end{align*}
with $\nu=20$.
For $m_i$, we considered the following two scenarios:
$$
{\rm (A)} \ \ m_i=\exp(-0.5+0.5x_{i1}+x_{i2}), \ \ \ \ \ \ \ 
{\rm (B)} \ \ m_i=\exp(|x_{i1}|x_{i2}).
$$
Since we use the two covariates $(x_{i1}, x_{i2})$, the Poisson-gamma model is correctly specified in scenario (A), but is misspecified in scenario (B).  
Here $n_i$ is a known constant and is set in the same way as Section~\ref{sec:ex-FH}.
The observed Poisson rate is $y_i=z_i/n_i$, and the benchmark target is $C = m^{-1}\sum_{i=1}^m \lambda_i^{\ast}$ as external benchmarking, where $\lambda_i^{\ast}$ is the true value.

We fitted the Poisson gamma model (\ref{PG}) to the synthetic data by generating 8000 posterior draws after discarding the first 2000 samples as the burn-in period. 
Based on the posterior draws, we computed the hierarchical Bayes (HB) estimator of $\lambda_i$ given by the posterior mean, and the proposed two benchmarked estimators by HET and EET. 
Furthermore, we adopted the benchmarked Bayes method via posterior projection with the benchmarked estimators of the form, $\widehat{\theta}_i^\mathrm{CB} = \widehat{\theta}_i + w_i(C - \sum_{i=1}^m w_i\widehat{\theta}_i)/s \phi_i$ where $s=\sum_{i=1}^mw_i^2\phi_i$, $\phi_i=w_i$ for PP1, and $\phi_i=w_i/\widehat{\theta}_i$ for PP2.

Based on 2000 Monte Carlo replications, mean squared errors (MSE) of point estimates and coverage probability (CP) and average length (AL) of $95\%$ credible intervals are calculated, which are reported in Table~\ref{tab:sim-PG}.
In Scenario (A), where the mean structure is correctly specified, HB tends to achieve the lowest MSE, while HET resulted in slightly higher in cases (I) and (II). 
In contrast, when the mean is misspecified as in Scenario (B) HET yielded the lowest MSE across all distributional settings. 
EET and PP1/PP2 consistently resulted in higher MSEs, especially in Scenario (III).

The table also shows that HB and HET both maintain nominal coverage 95\%.
In contrast, CPs for EET and PP2 declined, especially when the degree of model misspecification is severe, as in (B) and (II)-(III). 
PP1 generally exhibits greater under-coverage. 
With respect to AL, in Scenario (A), HB produced the narrowest credible intervals, followed by EET, while HET and PP2 resulted in wider intervals. 
Although ALs for PP1 are similar to those for HB, the poor coverage of PP1 indicates that these intervals for PP1 are too narrow.  
In Scenario (B), both HET and PP2 continued to produce wider credible intervals than HB, EET, and PP1, reflecting more conservative inference under model misspecification. 
However, since the HET attains smaller MSEs than PP2 in all cases, the result highlights the superiority of the proposed HET. 
In this regard, PP1 performed poorly in all cases of Scenario (B), with large MSEs and overly tight credible intervals, leading to undercoverage. 

Overall, the results demonstrate the robustness of the proposed HET under model misspecification. 
It performed competitively in Scenario (A) and showed superiority in Scenario (B). 
By contrast, PP is less reliable, consistently yielding higher MSEs, with PP1 proving especially problematic across all cases considered in this simulation study. 

The patterns observed in the first and second simulations differ. 
This may be partly due to differences not only in the misspecification of the mean structure but also in the degree of distributional misspecification; 
while the distributions of $v_i$ are all symmetric, the shapes of $\lambda_i$ vary across scenarios.

\begin{table}[tb]
\caption{
Mean squared errors (MSE) of posterior means, coverage probability (CP) and average length (AL) of $95\%$ credible intervals obtained by unconstrained HB posterior as well as the benchmarked posterior by hierarchical Bayesian entropic tilting (HET), empirical Bayesian entropic tilting (EET), and two types of posterior projection (PP1 and PP2), for Poisson-gamma models. 
The results are based on 2000 Monte Carlo replications.  
\label{tab:sim-PG}
}
\begin{center}
{
\begin{tabular}{ccccccccccccccc}
\hline
&& &\multicolumn{3}{c}{MSE} && \multicolumn{3}{c}{CP (\%)} && \multicolumn{3}{c}{AL} \\
DGP & Methods&  & (I) & (II) & (III) &  & (I) & (II) & (III) &  & (I) & (II) & (III) \\
\hline
 & HB &  & 0.053 & 0.049 & 0.137 &  & 96.2 & 96.7 & 94.5 &  & 0.874 & 0.867 & 1.139 \\
 & HET &  & 0.056 & 0.052 & 0.134 &  & 96.0 & 96.5 & 94.2 &  & 0.904 & 0.895 & 1.175 \\
(A) & EET &  & 0.058 & 0.054 & 0.150 &  & 94.2 & 94.8 & 92.0 &  & 0.850 & 0.840 & 1.165 \\
 & PP1 &  & 0.056 & 0.053 & 0.146 &  & 93.0 & 93.5 & 86.8 &  & 0.885 & 0.878 & 1.131 \\
 & PP2 &  & 0.061 & 0.057 & 0.149 &  & 95.6 & 96.1 & 92.6 &  & 0.919 & 0.910 & 1.182 \\
 \hline
 & HB &  & 0.171 & 0.174 & 0.213 &  & 94.0 & 93.9 & 94.6 &  & 1.341 & 1.336 & 1.490 \\
 & HET &  & 0.169 & 0.171 & 0.210 &  & 93.8 & 93.6 & 94.3 &  & 1.401 & 1.396 & 1.540 \\
(B) & EET &  & 0.180 & 0.182 & 0.266 &  & 91.8 & 91.6 & 91.5 &  & 1.391 & 1.384 & 1.581 \\
 & PP1 &  & 0.181 & 0.183 & 0.247 &  & 88.9 & 88.7 & 84.2 &  & 1.323 & 1.318 & 1.477 \\
 & PP2 &  & 0.177 & 0.179 & 0.276 &  & 92.4 & 92.2 & 91.2 &  & 1.414 & 1.409 & 1.581 \\
\hline
\end{tabular}
}
\end{center}
\end{table}

\section{Application: estimation of mortality rate in Japanese prefectures}

We demonstrate the proposed method through an application to the standardized mortality ratio (SMR) for cerebrovascular diseases among men in $m=47$ prefectures of Japan. 
Using data from the National Survey of Family Income and Expenditure, we include the prefecture-level health care expenditure and number of healthcare workers as covariates.
Let $n_i \ (i=1,\ldots,m)$ denote the expected number of mortality, $z_i$ denote the observed number of mortality, and $\bx_i=(x_{i1}, x_{i2})$ denote the vector of covariates.
Then, the direct estimate of SMR is defined as $y_i=z_i/n_i$.
In the dataset, the expected mortality $n_i$ ranges from $14.3$ to $245.8$, and the sampling ${\rm Var}(y_i)\approx y_i/n_i$ ranges from $0.38$ to $7.88$. 

For small area estimation of SMR, we fit the Poisson-gamma model 
$$
n_iy_i\sim {\rm Po}(n_i\lambda_i), \ \ \ \ \lambda_i\sim {\rm Ga}(\nu m_i, \nu), \ \ \ \  \log m_i=\beta_0+\beta_1  x_{i1} + \beta_2x_{i2},
$$
to estimate the true mortality rate $\lambda_i$, as it is typically adopted in the literature \citep[e.g.][]{kubokawa2014measuring}.
We specified the prior distributions as $\beta_0, \beta_1,\beta_2\sim N(0, 10^3)$ and $\nu\sim {\rm Ga}(1,1)$.
We generated 30000 posterior draws after discarding the first 20000 draws as the burn-in period, which required 1.94 seconds of computation time. 
Convergence of the MCMC chains was assessed using the Geweke diagnostic 
\citep{geweke1992evaluating}, with all absolute values of the Geweke $z$-scores below~2, indicating that the chains had reached stationarity.
The posterior means (standard deviations) of the parameters are as follows: 
$0.68\ (0.14)$ for $\beta_0$, $4.64\ (0.02)$ for $\beta_1$, $-0.05\ (0.02)$ for $\beta_2$ and $0.03\ (0.02)$ for $\nu$.

We consider the mean constraint given by $\sum_{i=1}^m w_i \hat{\lambda}_i=\sum_{i=1}^m w_i y_i$ with $w_i=n_i/\sum_{i'=1}^m n_{i'}$, which is typically required for the relevance of the policy.
Based on the posterior samples, we applied the hierarchical Bayes entropic tilting (HET) by using Algorithm~1 and got $\gamma=0.086$, which gives the tilted posterior of $\lambda_i$ and benchmarked estimates as the posterior means. 
For comparison, we also applied the posterior projection (PP) method with two benchmarked estimates (PP1 and PP2), as described in Section~\ref{sec:PG-sim}.
The computation times for the three methods were $0.073$ seconds for HET, $0.033$ seconds for PP1, and $0.050$ seconds for PP2.

\begin{figure}[htb!]
\centering
\includegraphics[width=\textwidth,clip]{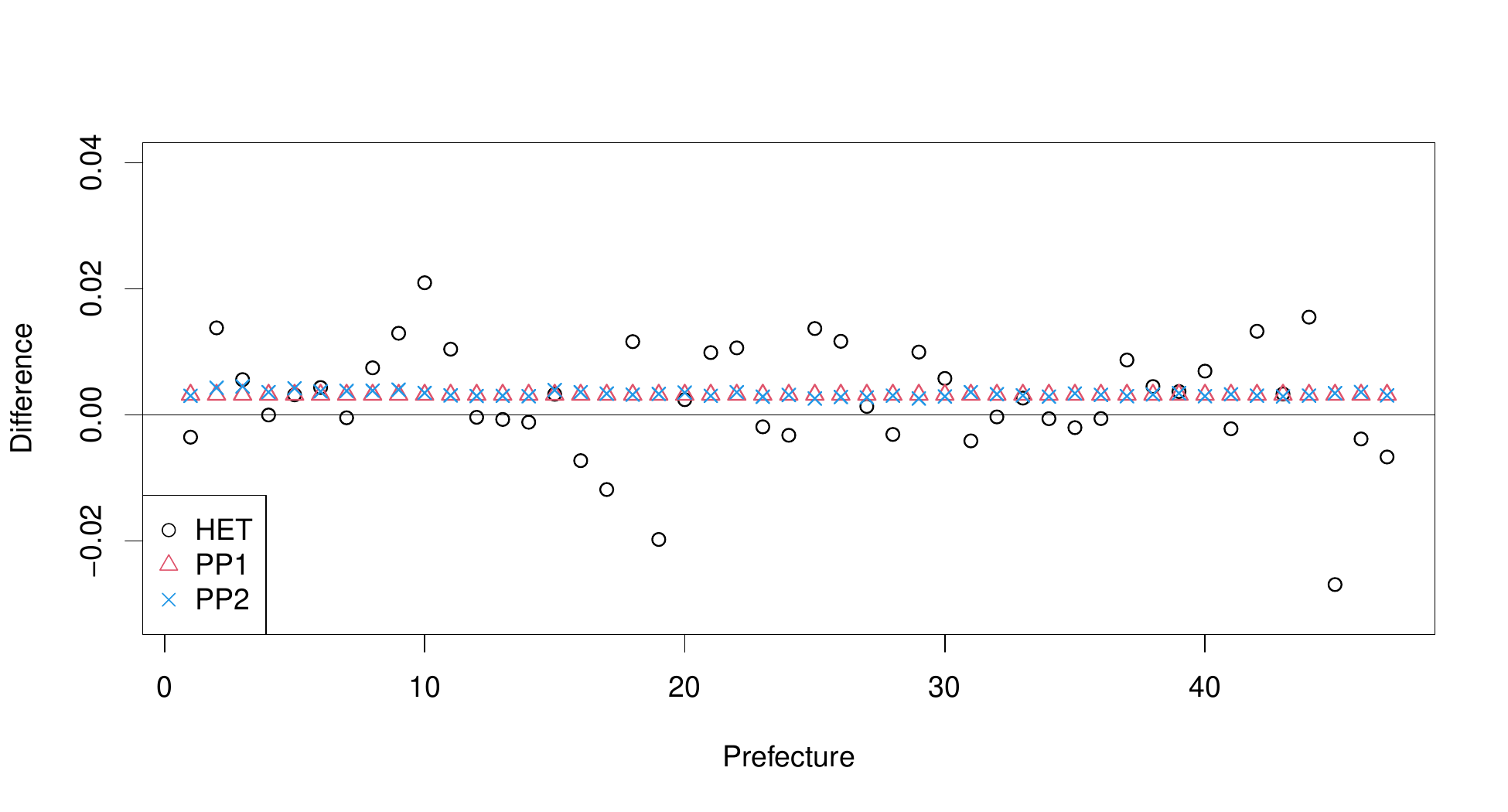}
\caption{The differences between the benchmarked estimates (ET, WL1 and WL2) and the original posterior means obtained by the Poisson-gamma model.
\label{fig:app}
}
\end{figure}

The differences between the benchmarked estimates and the original posterior means obtained by the Poisson–gamma model are shown in Figure~\ref{fig:app}. 
It is observed that both PP1 and PP2 provide uniform adjustments across areas, whereas ET applies area-specific modifications that vary across posterior samples. 
This reflects the fact that PP1 and PP2 are based on simple benchmarked point estimators, while ET adjusts the entire posterior distribution to satisfy the  the constraint, thereby incorporating posterior variances and other sources of uncertainty that PP1 and PP2 cannot take into account. 
It is therefore natural to regard the results of ET as preferable to those of posterior projection, since the latter constitutes an essentially ad-hoc adjustment without a solid theoretical foundation, as shown to perform worse than ET in the simulation study in Section~4.   

We also computed $95\%$ credible intervals from the original and benchmarked posteriors.
The average interval lengths was $6.32$ for both the original posterior and HET, while PP1 and PP2 produced shorter average interval lengths of $6.26$. 
This indicates that the proposed HET preserves the uncertainty quantification of the original posterior, whereas PP1 and PP2 lead to marginally overconfident results, consistent with the findings in Section~4.

\section{Concluding remarks}

The benchmark estimator by leveraging the entropic tilting provides both reasonable point estimates and uncertainty measures through tilted posterior distributions. 
We have illustrated the entropic tilting approach using representative models in small area estimation.
Importantly, the proposed method is broadly applicable to more general small area estimation problems, such as unit-level models \citep[e.g.][]{molina2014small} or models with estimated sampling variance \citep[e.g.][]{sugasawa2017bayesian}, 
provided that hierarchical posteriors of small area parameters can be obtained. 
We also note that our real data illustration is necessarily limited in scope, focusing on the performance of ET relative to a few existing approaches.  
A more extensive application to large-scale survey data would offer a clearer demonstration of the benefits of the method, which we leave for future study.


\vspace{0.5cm}
\appendix 
\begin{center}
{\bf {\large Appendix A1: Proof of Theorem~1}}
\end{center}

It holds that $\exp\{ w_i\bga^\top \bh(\theta_i)\}=1+w_i^{\ast} \bga^\top\bh(\theta_i)$ for some $w_i^{\ast}\in (0, w_i)$, noting that $w_i^{\ast}=O(m^{-1})$ since $w_i=O(m^{-1})$.
We then have 
\begin{align*}
V_m
&=\int \left[\prod_{i=1}^m \Big\{1+w_i^{\ast}\bga^\top E[\bh(\theta_i)|\bta, \by]\Big\}\right] \pi(\bta|\by)d\bta\\
&=\int \left\{1+\sum_{i=1}^m w_i^{\ast} \bga^\top E[\bh(\theta_i)|\bta,\by] +r(\gamma, \bta, \by)\right\}\pi(\bta|\by)d\bta,
\end{align*}
where 
$$
r(\gamma, \bta, \by)=\sum_{l=2}^m \sum_{(i_1,\ldots,i_l)\subset(1,\ldots,m)}
w_{i_1}^{\ast}\cdots w_{i_l}^{\ast} \bga^\top E[\bh(\theta_{i_1})] \cdots \bga^\top E[\bh(\theta_{i_l})].
$$
Since $w_{i_1}^{\ast}\cdots w_{i_l}^{\ast}=O(m^{-l})$ and the number of terms in the summation over $(i_1,\ldots,i_l)\subset(1,\ldots,m)$ is at most $O(m^{l})$, so that $r(\gamma, \bta, \by)=O(1)$. 
Hence, it holds that $V_m=1+\sum_{i=1}^m w_i^{\ast} \bga^\top E[\bh(\theta_i)]+r(\gamma, \by)$ with $r(\gamma,\by)=\int r(\gamma, \bta, \by) \pi(\bta|\by)d\bta$, which means that $V_m$ is a $m$th order polynomial function of $\gamma$ with coefficients being $O(1)$.  
On the other hand, 
\begin{align*}
&V_{m(i)}(\bga)\\
&\equiv \int \bh(\theta_i)\prod_{j=1}^m \exp\{ w_i\bga^\top \bh(\theta_i) \}f(\theta_j|\bta, \by)\pi(\bta|\by)d\bta\\
&=\int \Big\{E[\bh(\theta_i)|\bta,\by]+w_i^{\ast}  E[\{\bga^\top\bh(\theta_i)\}\bh(\theta_i) |\bta, \by]\Big\}  
\prod_{j=1, j\neq i}^m \Big\{1+\gamma w_j^{\ast} \bga^\top E[\bh(\theta_i)|\bta,\by]\Big\} \pi(\bta|\by)d\bta \\
&=\int 
\frac{ E[\bh(\theta_i)|\bta,\by]+w_i^{\ast} E[\{\bga^\top\bh(\theta_i)\}\bh(\theta_i) |\bta, \by] }
{1+\gamma w_i^{\ast} \bga^\top E[\bh(\theta_i)|\bta,\by]} 
\prod_{j=1}^m \Big\{1+\gamma w_j^{\ast} \bga^\top E[\bh(\theta_j)|\bta,\by]\Big\} \pi(\bta|\by)d\bta \\
&=\int \Big\{ E[\bh(\theta_i)|\bta,\by]+O(m^{-1}) \Big\} \left\{1+ \sum_{i=1}^m w_i^{\ast} \bga^\top E[\bh(\theta_i)|\bta],\by +r(\bga, \bta,\by)\right\} \pi(\bta|\by)d\bta\\
&= E[\bh(\theta_i)|\by] +  \sum_{i=1}^m w_i^{\ast}\int E[\bh(\theta_i)|\bta,\by] E[\bga^\top \bh(\theta_i)|\bta,\by]\pi(\bta|\by)d\bta\\
& \ \ \ \ \ \ \ \ \ 
+\int E[\bh(\theta_i)|\bta,\by] r(\bga, \bta, \by)\pi(\bta|\by)d\bta,
\end{align*}
indicating that $V_{m(i)}(\bga)$ is a $m$th order polynomial function of $\bga$ with coefficients being $O(1)$. 
Here the equation for $\bga$ can be expressed as $\sum_{i=1}^m w_i V_{m(i)}(\bga)/V_m(\bga)= C$, which is equivalent to $\sum_{i=1}^m w_i V_{m(i)}(\bga)= C \cdot V_m(\bga)$.
Both sides are $m$th order polynomial function of $\bga$ with coefficients being $O(1)$ since $w_i=O(m^{-1})$, whereby we have $\bga=O(1)$.

Noting that $g(\bth|\by)/f(\bth|\by)= V_m(\gamma)^{-1}\prod_{i=1}^m \exp\{\bga^\top \bh(\theta_i)\}$, the Kullbuck-Leibler distance between $g(\bth|\by)$ and $f(\bth|\by)$ is given by
\begin{align*}
{\rm KL}_{g|f}=\int \log\left(\frac{f(\bth|\by)}{g(\bth|\by)}\right)f(\bth|\by)d\bth =
\log V_m(\gamma) - \sum_{i=1}^m w_i \bga^\top E_f[\bh(\theta_i)],
\end{align*}
where $E_f$ denotes the expectation with respect to $f(\bth|\by)$.
Using the asymptotic expansion of the posterior expectations \citep{tierney1986accurate}, it holds that 
\begin{align*}
\log V_m&=\log \left[\prod_{i=1}^m \int\exp\{\bga^\top \bh(\theta_i)\}f(\theta_i|\btah)d\theta_i+O(m^{-2})\right]\\
&=\sum_{i=1}^m \log \int\exp\{\bga^\top \bh(\theta_i)\}f(\theta_i|\btah)d\theta_i + O(m^{-2}),
\end{align*}
where $\btah$ is the posterior mode of $\pi(\bta|\by)$.
Since $w_i=O(m^{-1})$, we expand $\exp\{\bga^\top \bh(\theta_i)\}$ to get 
\begin{align*}
\log V_m&=\sum_{i=1}^m \log \left\{1 + w_i  \bga^\top \Eh[\bh(\theta_i)]+\frac{w_i^2}{2}\bga^\top \Eh[\bh(\theta_i)\bh(\theta_i)^\top]\bga +O(m^{-3})\right\}\\
&=\sum_{i=1}^m w_i  \bga^\top \Eh[\bh(\theta_i)] + \frac12\sum_{i=1}^m w_i^2\bga^\top \Eh[\bh(\theta_i)\bh(\theta_i)^\top] \bga +O(m^{-2}),
\end{align*}
where $\Eh$ denotes the expectation with respect to $f(\theta_i|\btah, \by)$.
Then, we have 
\begin{align*}
{\rm KL}_{g|f}=\sum_{i=1}^mw_i \bga^\top \big\{\Eh[\bh(\theta_i)]-E_f[\bh(\theta_i)]\big\}+\frac12 \sum_{i=1}^m w_i^2\bga^\top \Eh[\bh(\theta_i)\bh(\theta_i)^\top] \bga+O(m^{-2}),
\end{align*}
which is $O(m^{-1})$ since $\Eh[\bh(\theta_i)]-E_f[\bh(\theta_i)]=O(m^{-1})$ and $\bga=O(1)$.

\vspace{5mm}
\begin{center}
{\bf {\large Appendix A2: MCMC algorithms}}
\end{center}

The Gibbs sampler of the Fay-Herriot model to obtain an MCMC sample from the posterior $\pi(\bu, \bta|\by)$ is described as follows: 
\begin{itemize}
\item[-] 
\textbf{Sampling $\theta_i$:}
For $i=1,\ldots,m$, the full conditional of $\theta_i$ is $N(\tht_i(u_i,\bta), \sit_i^2(u_i,\bta))$, where $\tht_i(u_i,\bta)$ and $\sit_i^2(u_i,\bta)$ are given in (\ref{FH-pos}).

\item[-] 
\textbf{Sampling $A$:}
Assuming the prior $A\sim {\rm IG}(n_0,s_0)$, the full conditional of $A$ is given by ${\rm IG}(n_0+m/2, s_0+\sum_{i=1}^m(\theta_i-\bx_i^\top\bbe)^2/2u_i)$

\item[-] 
\textbf{Sampling $\bbe$:}
Assuming the prior $\bbe\sim N(\bb_0,\bB_0)$, the full conditional of $\bbe$ is given by $N(\bB_\beta^{-1} \bb_\beta, \bB_\beta^{-1})$ where $\bB_{\beta}=A^{-1}\bX^\top \bU^{-1}\bX+\bB_0^{-1}$ and $\bb_\beta=A^{-1}\bX^\top \bU^{-1}\bth$ with $\bU={\rm diag}(u_1,\ldots,u_m)$.

\item[-]
\textbf{Sampling $u_i$:}
For $i=1,\ldots,m$, the full conditional density of $u_i$ is proportional to $\pi(u_i;\lambda)u_i^{-1/2}\exp\{-(\theta_i-\bx_i^\top\bbe)/2Au_i\}$. 
For some priors (e.g. Laplace and horseshoe), the full conditional is a familiar distribution.
\end{itemize}

Also, the Gibbs sampler to obtain an MCMC sample from the posterior $\pi(\bta|\by)$ under the Poisson-gamma model is given as follows: 
\begin{itemize}
\item[-] 
\textbf{Sampling $\lambda_i$:}
The full conditional of $\la_i$ is ${\rm Ga}(n_iy_i+\nu m_i, n_i+\nu)$. 

\item[-]
\textbf{Sampling $\bbe$ and $\nu$:}
The joint full conditional of $\bbe$ and $\nu$ has the density proportional to
\[
\pi(\bbe,\nu)\prod_{i=1}^m p_{\text{Ga}}\left(\la_i;\nu \exp(\bx_i^\top\bbe),\nu\right),
\]
where $\pi(\bbe,\nu)$ denotes the prior density of $(\bbe,\nu)$. 
It is possible to sample $\bbe$ and $\nu$ alternately by using the Metropolis-Hastings algorithm. 
\end{itemize}

\bigskip
\section*{Data Availability Statement}
Codes for simulation experiment in this study are publicly available on the authors’ GitHub page (URL will be provided upon acceptance). 
However, the SMR data analyzed in this paper cannot be publicly shared due to data ownership and usage restrictions beyond the authors’ control.

\vspace{0.3cm}
\bibliographystyle{chicago}
\bibliography{ref}

\begin{thebibliography}{}

\bibitem[\protect\citeauthoryear{Battese, Harter, and Fuller}{Battese et~al.}{1988}]{battese88}
Battese, G.~E., R.~M. Harter, and W.~A. Fuller (1988).
\newblock An error-components model for prediction of county crop areas using survey and satellite data.
\newblock {\em Journal of the American Statistical Association\/}~{\em 83}, 28--36.

\bibitem[\protect\citeauthoryear{Bell, Datta, and Ghosh}{Bell et~al.}{2013}]{bell2013benchmarking}
Bell, W.~R., G.~S. Datta, and M.~Ghosh (2013).
\newblock Benchmarking small area estimators.
\newblock {\em Biometrika\/}~{\em 100\/}(1), 189--202.

\bibitem[\protect\citeauthoryear{Carvalho, Polson, and Scott}{Carvalho et~al.}{2010}]{carvalho2010horseshoe}
Carvalho, C.~M., N.~G. Polson, and J.~G. Scott (2010).
\newblock The horseshoe estimator for sparse signals.
\newblock {\em Biometrika\/}~{\em 97\/}(2), 465--480.

\bibitem[\protect\citeauthoryear{Chikamatsu and Kubokawa}{Chikamatsu and Kubokawa}{2023}]{chikamatsu2023benchmarked}
Chikamatsu, K. and T.~Kubokawa (2023).
\newblock Benchmarked linear shrinkage prediction in the {F}ay--{H}erriot small area model.
\newblock {\em Scandinavian Journal of Statistics\/}~{\em 50\/}(2), 572--588.

\bibitem[\protect\citeauthoryear{Clayton and Kaldor}{Clayton and Kaldor}{1987}]{clayton1987empirical}
Clayton, D. and J.~Kaldor (1987).
\newblock Empirical bayes estimates of age-standardized relative risks for use in disease mapping.
\newblock {\em Biometrics\/}, 671--681.

\bibitem[\protect\citeauthoryear{Csisz{\'a}r}{Csisz{\'a}r}{1975}]{csiszar1975divergence}
Csisz{\'a}r, I. (1975).
\newblock I-divergence geometry of probability distributions and minimization problems.
\newblock {\em The annals of probability\/}, 146--158.

\bibitem[\protect\citeauthoryear{Datta, Ghosh, Steorts, and Maples}{Datta et~al.}{2011}]{datta2011bayesian}
Datta, G., M.~Ghosh, R.~Steorts, and J.~Maples (2011).
\newblock Bayesian benchmarking with applications to small area estimation.
\newblock {\em Test\/}~{\em 20}, 574--588.

\bibitem[\protect\citeauthoryear{Datta and Mandal}{Datta and Mandal}{2015}]{datta2015small}
Datta, G.~S. and A.~Mandal (2015).
\newblock Small area estimation with uncertain random effects.
\newblock {\em Journal of the American Statistical Association\/}~{\em 110\/}(512), 1735--1744.

\bibitem[\protect\citeauthoryear{Fay and Herriot}{Fay and Herriot}{1979}]{fay1979estimates}
Fay, R.~E. and R.~A. Herriot (1979).
\newblock Estimates of income for small places: an application of {J}ames-{S}tein procedures to census data.
\newblock {\em Journal of the American Statistical Association\/}~{\em 74\/}(366a), 269--277.

\bibitem[\protect\citeauthoryear{Geweke}{Geweke}{1992}]{geweke1992evaluating}
Geweke, J. (1992).
\newblock Evaluating the accuracy of sampling-based approaches to the calculations of posterior moments.
\newblock {\em Bayesian statistics\/}~{\em 4}, 641--649.

\bibitem[\protect\citeauthoryear{Ghosh}{Ghosh}{1992}]{ghosh92cb}
Ghosh, M. (1992).
\newblock Constrained {B}ayes estimation with applications.
\newblock {\em Journal of the American Statistical Association\/}~{\em 87}, 533--540.

\bibitem[\protect\citeauthoryear{Ghosh, Kubokawa, and Kawakubo}{Ghosh et~al.}{2015}]{ghosh2015benchmarked}
Ghosh, M., T.~Kubokawa, and Y.~Kawakubo (2015).
\newblock Benchmarked empirical {B}ayes methods in multiplicative area-level models with risk evaluation.
\newblock {\em Biometrika\/}~{\em 102\/}(3), 647--659.

\bibitem[\protect\citeauthoryear{Ghosh and Maiti}{Ghosh and Maiti}{2004}]{ghosh2004small}
Ghosh, M. and T.~Maiti (2004).
\newblock Small-area estimation based on natural exponential family quadratic variance function models and survey weights.
\newblock {\em Biometrika\/}~{\em 91\/}(1), 95--112.

\bibitem[\protect\citeauthoryear{Ghosh and Maiti}{Ghosh and Maiti}{2008}]{ghosh2008empirical}
Ghosh, M. and T.~Maiti (2008).
\newblock Empirical {B}ayes confidence intervals for means of natural exponential family-quadratic variance function distributions with application to small area estimation.
\newblock {\em Scandinavian Journal of Statistics\/}~{\em 35\/}(3), 484--495.

\bibitem[\protect\citeauthoryear{Ghosh and Steorts}{Ghosh and Steorts}{2013}]{ghosh2013two}
Ghosh, M. and R.~C. Steorts (2013).
\newblock Two-stage benchmarking as applied to small area estimation.
\newblock {\em Test\/}~{\em 22}, 670--687.

\bibitem[\protect\citeauthoryear{Ghosh and Steorts}{Ghosh and Steorts}{2019}]{ghosh2019survey}
Ghosh, M. and R.~C. Steorts (2019).
\newblock Some variants of constrained estimation in finite population sampling.
\newblock {\em International Statistical Review\/}~{\em 87\/}(S1), S90--S103.

\bibitem[\protect\citeauthoryear{Janicki and Vesper}{Janicki and Vesper}{2017}]{janicki2017}
Janicki, R. and A.~Vesper (2017).
\newblock Benchmarking techniques for reconciling {B}ayesian small area models at distinct geographic levels.
\newblock {\em Statistical Methods \& Applications\/}~{\em 26}, 557--581.

\bibitem[\protect\citeauthoryear{Jaynes}{Jaynes}{1957}]{jaynes1957information}
Jaynes, E.~T. (1957).
\newblock Information theory and statistical mechanics.
\newblock {\em Physical review\/}~{\em 106\/}(4), 620.

\bibitem[\protect\citeauthoryear{Kubokawa, Hasukawa, and Takahashi}{Kubokawa et~al.}{2014}]{kubokawa2014measuring}
Kubokawa, T., M.~Hasukawa, and K.~Takahashi (2014).
\newblock On measuring uncertainty of benchmarked predictors with application to disease risk estimate.
\newblock {\em Scandinavian Journal of Statistics\/}~{\em 41\/}(2), 394--413.

\bibitem[\protect\citeauthoryear{Molina, Nandram, and Rao}{Molina et~al.}{2014}]{molina2014small}
Molina, I., B.~Nandram, and J.~Rao (2014).
\newblock Small area estimation of general parameters with application to poverty indicators: A hierarchical {B}ayes approach.
\newblock {\em The Annals of Applied Statistics\/}~{\em 8\/}(2), 852.

\bibitem[\protect\citeauthoryear{Okonek and Wakefield}{Okonek and Wakefield}{2022}]{okonek2022computationally}
Okonek, T. and J.~Wakefield (2022).
\newblock A computationally efficient approach to fully {B}ayesian benchmarking.
\newblock {\em arXiv preprint arXiv:2203.12195\/}.

\bibitem[\protect\citeauthoryear{Park and Casella}{Park and Casella}{2008}]{park2008bayesian}
Park, T. and G.~Casella (2008).
\newblock The {B}ayesian lasso.
\newblock {\em Journal of the American Statistical Association\/}, 681--686.

\bibitem[\protect\citeauthoryear{Pfeffermann and Tiller}{Pfeffermann and Tiller}{2006}]{pfeffermann2006small}
Pfeffermann, D. and R.~Tiller (2006).
\newblock Small-area estimation with state--space models subject to benchmark constraints.
\newblock {\em Journal of the American Statistical Association\/}~{\em 101\/}(476), 1387--1397.

\bibitem[\protect\citeauthoryear{Rao and Molina}{Rao and Molina}{2015}]{rao2015small}
Rao, J. N.~K. and I.~Molina (2015).
\newblock {\em Small area estimation}.
\newblock John Wiley \& Sons.

\bibitem[\protect\citeauthoryear{Steorts and Ghosh}{Steorts and Ghosh}{2013}]{steorts2013estimation}
Steorts, R.~C. and M.~Ghosh (2013).
\newblock On estimation of mean squared errors of benchmarked empirical {B}ayes estimators.
\newblock {\em Statistica Sinica\/}, 749--767.

\bibitem[\protect\citeauthoryear{Steorts, Schmid, and Tzavidis}{Steorts et~al.}{2020}]{steorts2020smoothing}
Steorts, R.~C., T.~Schmid, and N.~Tzavidis (2020).
\newblock Smoothing and benchmarking for small area estimation.
\newblock {\em International Statistical Review\/}~{\em 88\/}(3), 580--598.

\bibitem[\protect\citeauthoryear{Sugasawa, Kawakubo, and Ogasawara}{Sugasawa et~al.}{2020}]{sugasawa2020spatial}
Sugasawa, S., Y.~Kawakubo, and K.~Ogasawara (2020).
\newblock Small area estimation with spatially varying natural exponential families.
\newblock {\em Journal of Statistical Computation and Simulation\/}~{\em 90\/}(6), 1039--1056.

\bibitem[\protect\citeauthoryear{Sugasawa and Kubokawa}{Sugasawa and Kubokawa}{2020}]{sugasawa2020small}
Sugasawa, S. and T.~Kubokawa (2020).
\newblock Small area estimation with mixed models: a review.
\newblock {\em Japanese Journal of Statistics and Data Science\/}~{\em 3}, 693--720.

\bibitem[\protect\citeauthoryear{Sugasawa, Kubokawa, and Ogasawara}{Sugasawa et~al.}{2017}]{sugasawa2017empirical}
Sugasawa, S., T.~Kubokawa, and K.~Ogasawara (2017).
\newblock Empirical uncertain {B}ayes methods in area-level models.
\newblock {\em Scandinavian Journal of Statistics\/}~{\em 44\/}(3), 684--706.

\bibitem[\protect\citeauthoryear{Sugasawa, Tamae, and Kubokawa}{Sugasawa et~al.}{2017}]{sugasawa2017bayesian}
Sugasawa, S., H.~Tamae, and T.~Kubokawa (2017).
\newblock Bayesian estimators for small area models shrinking both means and variances.
\newblock {\em Scandinavian Journal of Statistics\/}~{\em 44\/}(1), 150--167.

\bibitem[\protect\citeauthoryear{Tallman and West}{Tallman and West}{2022}]{tallman2022entropic}
Tallman, E. and M.~West (2022).
\newblock On entropic tilting and predictive conditioning.
\newblock {\em arXiv preprint arXiv:2207.10013\/}.

\bibitem[\protect\citeauthoryear{Tang, Ghosh, Ha, and Sedransk}{Tang et~al.}{2018}]{tang2018modeling}
Tang, X., M.~Ghosh, N.~S. Ha, and J.~Sedransk (2018).
\newblock Modeling random effects using global--local shrinkage priors in small area estimation.
\newblock {\em Journal of the American Statistical Association\/}~{\em 113\/}(524), 1476--1489.

\bibitem[\protect\citeauthoryear{Tierney and Kadane}{Tierney and Kadane}{1986}]{tierney1986accurate}
Tierney, L. and J.~B. Kadane (1986).
\newblock Accurate approximations for posterior moments and marginal densities.
\newblock {\em Journal of the American Statistical Association\/}~{\em 81\/}(393), 82--86.

\bibitem[\protect\citeauthoryear{Wang, Fuller, and Qu}{Wang et~al.}{2008}]{wang2008small}
Wang, J., W.~A. Fuller, and Y.~Qu (2008).
\newblock Small area estimation under a restriction.
\newblock {\em Survey methodology\/}~{\em 34\/}(1), 29.

\bibitem[\protect\citeauthoryear{West}{West}{2023}]{west2023perspectives}
West, M. (2023).
\newblock Perspectives on constrained forecasting.
\newblock {\em Bayesian Analysis\/}~{\em 1\/}(1), 1--27.

\bibitem[\protect\citeauthoryear{You, Rao, and Dick}{You et~al.}{2004}]{you2004}
You, Y., J.~N.~K. Rao, and P.~Dick (2004).
\newblock Benchmarking hierarchical bayes small area estimators in the canadian census undercoverage estimation.
\newblock {\em Statistics in Transition\/}~{\em 6}, 631--640.

\bibitem[\protect\citeauthoryear{Zhang and Bryant}{Zhang and Bryant}{2020}]{zhang2020fully}
Zhang, J.~L. and J.~Bryant (2020).
\newblock Fully {B}ayesian benchmarking of small area estimation models.
\newblock {\em Journal of official statistics\/}~{\em 36\/}(1), 197--223.

\end{thebibliography}

\end{document}